\documentclass[journal=jpccck,manuscript=article,layout=onecolumn]{achemso}
\usepackage{chemformula} 
\usepackage[T1]{fontenc} 
\usepackage{hyperref}
\usepackage{comment}
\usepackage{amssymb}

\author{Felipe Hawthorne}
\affiliation{Department of Physics, Federal University of Parana,
               R. Evaristo F. Ferreira da Costa, 81530-015, Curitiba, Brazil}
\alsoaffiliation[Federal University of Parana]{Interdisciplinary Center for Science, Technology, and Innovation (CICTI), Federal University of Parana, Av. Cel. Francisco H. dos Santos, 81530-000, Curitiba, Brazil}
\author{Leandro Seixas}
    \affiliation{Instituto de F{\'\i}sica Te\'orica, Universidade Estadual Paulista, Rua Dr. Bento Teobaldo Ferraz, 271, 01140-070, S\~ao Paulo, SP, Brazil}
\author{James M. Almeida}
  \affiliation{Ilum School of Science, Brazilian Center for Research in Energy and Materials (CNPEM), Zip Code 13083-970, Campinas, Sao Paulo, Brazil.}

\author{Cristiano F. Woellner}
\email{woellner@ufpr.br}
\affiliation{Department of Physics, Federal University of Parana,
               R. Evaristo F. Ferreira da Costa, 81530-015, Curitiba, Brazil}
\alsoaffiliation[Federal University of Parana]{Interdisciplinary Center for Science, Technology, and Innovation (CICTI), Federal University of Parana, Av. Cel. Francisco H. dos Santos, 81530-000, Curitiba, Brazil}

\author{Raphael M. Tromer}
\email{raphael.tromer@unb.br}
\affiliation[IF UnB]
{University of Brasília, Institute of Physics, Brasília, Federal District, Brazil.}

\title[An \textsf{achemso} demo]
  {Interpretable Machine Learning of Nanoparticle Stability through Topological Layer Embeddings}

\begin{document}

\begin{abstract}
The stability of chemically complex nanoparticles is governed by an immense configurational space arising from heterogeneous local atomic environments across surface and interior regions. Efficiently identifying low-energy configurations within this space remains a central challenge for first-principles–based materials discovery, particularly when the available reference data are limited. Here, we introduce a data-efficient and physically interpretable machine-learning framework based on a fragmented, layer-resolved descriptor that explicitly decomposes nanoparticles into surface, intermediate, and core environments using a topology-driven definition. This representation preserves a compact and fixed feature dimensionality while retaining spatial resolution, enabling controlled emphasis on different regions of the nanoparticle through physically motivated weighting schemes. Coupled with gradient-boosted decision tree models and a ranking-based learning strategy, the proposed framework enables accurate identification of the most stable nanoparticle configurations using only a few hundred density functional theory reference calculations. Ranking performance metrics demonstrate near-saturation of correlation, high top-k recall, and rapidly vanishing regret at moderate training-set sizes, highlighting the strong data efficiency of the approach. Beyond predictive performance, layer-weighting and SHAP-based interpretability analyses reveal how surface segregation, coordination topology, and local chemical disorder contribute differently to stability across spatial regions of the nanoparticle. These insights provide a transparent physical interpretation of the learned models and establish a natural pathway toward active learning–driven exploration of complex nanoparticle configurational spaces.
\end{abstract}

\section{Introduction}

Understanding and predicting the stability of chemically complex nanoparticles remains a fundamental challenge in nanoscale materials science~\cite{Behler2011AtomcenteredSF,10.1063/1.1584074,Ceder2010}. Recent advances in machine learning have further accelerated this field by enabling data-driven exploration and design of nanomaterials across vast structural and compositional spaces, complementing first-principles approaches in regimes where exhaustive sampling is computationally prohibitive~\cite{delBosque2026,Diao2025}. Unlike bulk materials, nanoparticles exhibit pronounced heterogeneity in local atomic environments, arising from variations in coordination, chemical composition, and bonding across surface, subsurface, and core regions. These spatially distinct environments coexist within a single finite structure and give rise to a vast and rugged configurational energy landscape, even at fixed global stoichiometry.  

In multicomponent metallic nanoparticles, this complexity is further amplified by chemical disorder~\cite{Polukhin2024,Nagarjuna2025}, coordination heterogeneity~\cite{MoreiraDaSilva2022,Mahin2025}, and competing local bonding motifs that can differ markedly between surface and interior regions~\cite{doi:10.1021/acsnano.4c16417,Kar2023}. As a result, identifying low-energy configurations through direct first-principles sampling becomes rapidly intractable, as the number of distinct atomic arrangements grows combinatorially with system size and chemical diversity. This intrinsic configurational complexity establishes a critical bottleneck for rational nanoparticle design and motivates the development of efficient, physically informed strategies for exploring stability landscapes.

A wide variety of structural descriptors have been proposed to encode atomic configurations for machine-learning models~\cite{Lin2023,Caro2019,Inada2021}. Atom-centered representations, such as symmetry functions~\cite{Behler2011}, SOAP descriptors~\cite{Bartok2013,Willatt2018}, and graph-based embeddings~\cite{Batzner2022,Chang2025,D3CP02143B}, have proven highly successful for learning local properties and constructing interatomic potentials. However, these approaches are typically high-dimensional and computationally demanding, and they often require large training datasets to achieve robust generalization~\cite{10.1063/1.4966192,https://doi.org/10.1002/anie.201709686}. As a consequence, their direct application to the efficient exploration of nanoparticle stability landscapes remains challenging in data-limited regimes. Recent studies have emphasized that the choice of structural descriptors is often the dominant factor governing data efficiency, transferability, and interpretability in machine-learning models for materials, frequently outweighing the impact of the specific learning algorithm employed~\cite{Tromer2025,Dau2023}.  

Global descriptors provide an alternative strategy by summarizing entire nanoparticles or clusters through averaged geometric~\cite{Inada2021}, topological~\cite{Galanakis2024}, or chemical features~\cite{Bartel2018}. While these representations can improve data efficiency, they do so at the cost of spatial resolution~\cite{Himanen2020}, effectively mixing information from surface, subsurface, and core environments. In practice, this loss of spatial discrimination makes it difficult to disentangle region-specific contributions to stability and to rationalize how surface-driven effects compete with bulk-like energetic trends~\cite{doi:10.1021/cr040090g,Bartok2013}.

For nanoparticles, and particularly for chemically complex and core--shell–like systems, this lack of spatial resolution becomes especially limiting. Surface atoms typically dominate reactivity and often play a decisive role in stability, while interior atoms largely control cohesive energy, elastic response, and structural rigidity~\cite{RevModPhys.77.371}. In multicomponent alloys, variations in chemical short-range order, coordination topology, and local bonding motifs can differ markedly between surface and core regions, even when the global composition is fixed~\cite{doi:10.1021/cr040090g}.  

As a consequence, stability cannot be reliably rationalized in terms of either purely surface-driven or purely bulk-like descriptors. Instead, it emerges from a coupled interplay between distinct spatial regions, whose relative energetic importance may vary across configurational space. In this context, there is growing recognition that physically interpretable machine-learning models are essential not only for predictive accuracy, but also for uncovering mechanistic insights and guiding rational materials design~\cite{Oviedo2022ExplainableML}. Current descriptor frameworks rarely provide a systematic and interpretable way to isolate these regional contributions while maintaining a compact representation suitable for data-efficient learning, limiting their applicability for rational screening and design of complex nanoparticles~\cite{Bartok2013}.

In this work, we introduce a fragmented, layer-resolved descriptor framework that explicitly decomposes nanoparticles into topological shells defined by their connectivity distance from the surface. By construction, this approach provides a physically motivated separation between surface, intermediate, and core environments without relying on arbitrary geometric cutoffs. The resulting representation preserves a compact and fixed feature dimensionality, independent of nanoparticle size or shape, while retaining essential spatial resolution across distinct regions of the structure.  

This layered embedding enables controlled emphasis on different parts of the nanoparticle through physically motivated weighting schemes, allowing one to systematically probe how surface, subsurface, and core environments contribute to stability. The descriptor is invariant to atom indexing, robust to small structural distortions, and directly applicable to nanoparticles with complex morphologies and chemical disorder. Importantly, it bridges the gap between fully local atom-centered descriptors and purely global representations by embedding spatial heterogeneity in a transparent, tunable, and interpretable manner.

To directly address the practical objective of identifying the most stable nanoparticle configurations with minimal computational effort, we formulate the learning task as a ranking problem rather than a strict energy regression. This perspective aligns naturally with high-throughput screening and active learning workflows, where the primary goal is to reliably down-select a small subset of low-energy candidates for expensive first-principles validation. Ranking-based formulations have been increasingly adopted in materials discovery workflows, particularly in active learning and high-throughput screening contexts, where relative ordering of candidates is more relevant than absolute property prediction~\cite{Himanen2020,Ma2025}.  

Within this framework, we employ gradient-boosted decision tree models~\cite{Chen2016}, which are particularly well suited for data-limited regimes and structured, physically informed descriptors. While neural-network architectures can achieve high accuracy when trained on very large datasets, gradient boosting offers a favorable balance between expressive power, robustness against overfitting, and computational efficiency when only a few hundred reference calculations are available~\cite{Chen2016,Friedman2001}.  

Finally, the combination of layer-weighted embeddings with post hoc interpretability tools enables a quantitative assessment of how different spatial regions of the nanoparticle govern energetic stability. By linking model predictions to physically meaningful descriptors associated with surface, intermediate, and core environments, the proposed approach provides insights that go beyond black-box prediction and establishes a scalable, interpretable framework for accelerated discovery and analysis of stable nanostructures in complex multicomponent systems.

\section{Methodology}

First-principles calculations were carried out within the framework of Density Functional Theory (DFT) as implemented in the \textsc{SIESTA}~\cite{Soler2002} code. The metallic nanoparticle was modeled in a cubic supercell with dimensions of $35 \times 35 \times 35$~\AA, sufficiently large to prevent interactions between periodic images. Brillouin zone sampling was performed using a Monkhorst--Pack grid at $\Gamma$ point.

Electron--ion interactions were described using norm-conserving pseudopotentials, and the electronic wave functions were expanded in a localized atomic orbital basis set of double-zeta with polarization (DZP)~\cite{SanchezPortal1996}. Exchange--correlation effects were treated within the generalized gradient approximation (GGA) using the Perdew--Burke--Ernzerhof (PBE) functional. A real-space mesh cutoff of 250~Ry was employed.

The Kohn--Sham equations were solved by direct diagonalization. Self-consistency was achieved using density matrix mixing with a weight of 0.035 and a convergence tolerance of $10^{-4}$. Spin polarization was not considered.

Structural optimization was performed using the conjugate gradient method while keeping the supercell volume fixed. Atomic positions were relaxed until the maximum force on each atom was below 0.02~eV/\AA, with a maximum atomic displacement of 0.1~\AA\ per relaxation step.

Atomic structures were analyzed through a graph-theoretical framework constructed directly from their atomic coordinates~\cite{Balaban1976}. Each structure consists of a set of $N$ atoms with positions $\{\mathbf{r}_i\}_{i=1}^{N}$ and chemical species $\{s_i\}_{i=1}^{N}$. A chemical connectivity graph $G = (V,E)$ was defined, where each vertex $i \in V$ corresponds to an atom and an undirected edge $(i,j) \in E$ exists if the interatomic distance $d_{ij}$ satisfies
\begin{equation}
d_{ij} \leq \lambda \, (r_i^{\mathrm{nat}} + r_j^{\mathrm{nat}}),
\end{equation}
where $r^{\mathrm{nat}}$ are ASE~\cite{ase-paper} natural cutoff radii and $\lambda$ is a global multiplicative factor. Distances were evaluated using the minimum image convention when periodic boundary conditions were present. This procedure provides an element-aware yet parameter-light definition of nearest-neighbor connectivity applicable to both crystalline and disordered systems.

For each atom $i$, the coordination number was defined as the degree of the corresponding graph node,
\begin{equation}
k_i = \sum_{j} A_{ij},
\end{equation}
where $A_{ij}$ is the adjacency matrix of $G$. An ideal bulk coordination number $k_{\mathrm{bulk}}$ was inferred automatically as the statistical mode of the $\{k_i\}$ distribution,
\begin{equation}
k_{\mathrm{bulk}} = \mathrm{mode}\left(\{k_i\}\right),
\end{equation}
under the assumption that bulk-like environments dominate the atomic population. Atoms with $k_i < k_{\mathrm{bulk}}$ were classified as surface seeds, reflecting the physical reduction of coordination at free surfaces and defects.

To define surface and bulk regions in a robust and topology-driven manner, a breadth-first search~\cite{Moore:1959} was performed on $G$ starting simultaneously from all surface seeds. This procedure assigns to each atom a topological distance $\ell_i$, defined as the minimum number of graph edges connecting atom $i$ to any surface seed. Formally,
\begin{equation}
\ell_i = \min_{j \in \mathcal{S}} \mathrm{dist}_{G(i,j)},
\end{equation}
where $\mathcal{S}$ denotes the set of surface seeds and $\mathrm{dist}_G$ is the shortest-path distance on $G$. Atoms satisfying $\ell_i < L$ were classified as surface atoms, while atoms with $\ell_i \geq L$ were assigned to the bulk, where $L$ is a user-defined number of topological layers. This definition naturally includes subsurface atoms and is invariant with respect to geometric distortions, surface roughness, and local disorder.

For any atomic subset $\Omega \subseteq V$ (total system, bulk, or surface), region-resolved geometric descriptors were computed. The number of atoms in the region is $N_\Omega = |\Omega|$. The mean coordination number and its standard deviation are given by
\begin{equation}
\langle k \rangle_\Omega = \frac{1}{N_\Omega} \sum_{i \in \Omega} k_i,
\end{equation}
\begin{equation}
\qquad
\sigma_k^\Omega = \sqrt{\frac{1}{N_\Omega} \sum_{i \in \Omega} (k_i - \langle k \rangle_\Omega)^2}.
\end{equation}
Bond-length statistics were computed using only edges fully contained within the region,
\begin{equation}
\mathcal{E}_\Omega = \{(i,j) \in E \mid i \in \Omega,\, j \in \Omega\},
\end{equation}
yielding the mean bond length $\langle d \rangle_\Omega$ and its standard deviation $\sigma_d^\Omega$.

Medium-range topological order was optionally characterized through a cycle analysis of $G$. For each edge $(i,j)$, the shortest alternative path between $i$ and $j$ excluding that edge was determined. When such a path of length $m-1$ exists, a cycle of size $m$ is defined. Cycles with sizes within a predefined interval $[m_{\min}, m_{\max}]$ were collected and mapped onto a canonical representation to remove rotational and mirror degeneracies. Let $n_m$ denote the number of distinct cycles of size $m$. The normalized ring fractions are defined as~\cite{LEROUX201070}
\begin{equation}
p_m = \frac{n_m}{\sum_{m} n_m},
\end{equation}
and a topological Shannon entropy was computed as
\begin{equation}
S_{\mathrm{topo}} = - \sum_{m} p_m \ln p_m,
\end{equation}
providing a scalar measure of network complexity and medium-range disorder.

Chemical order was quantified through region-resolved compositional and short-range order descriptors. The atomic fraction of species $A$ in region $\Omega$ is
\begin{equation}
c_A^\Omega = \frac{1}{N_\Omega} \sum_{i \in \Omega} \delta_{s_i,A},
\end{equation}
where $\delta$ is the Kronecker delta. For each ordered pair of chemical species $(A,B)$, the neighbor fraction $P_{AB}^\Omega$ was defined as
\begin{equation}
P_{AB}^\Omega =
\frac{\sum_{i \in \Omega} \sum_{j \in \mathcal{N}_i^\Omega} \delta_{s_i,A} \, \delta_{s_j,B}}
{\sum_{i \in \Omega} \delta_{s_i,A} \, |\mathcal{N}_i^\Omega|},
\end{equation}
where $\mathcal{N}_i^\Omega$ denotes the set of neighbors of atom $i$ restricted to region $\Omega$. Based on these quantities, the Warren--Cowley short-range order parameter~\cite{Cowley1950} was evaluated as
\begin{equation}
\alpha_{AB}^\Omega = 1 - \frac{P_{AB}^\Omega}{c_B^\Omega}.
\end{equation}
Negative values of $\alpha_{AB}^\Omega$ indicate a preference for $A$--$B$ nearest-neighbor pairs, while positive values indicate chemical avoidance or segregation.

Chemical disorder within each region was further quantified by a chemical Shannon entropy~\cite{Lederer2018,Shannon1948},
\begin{equation}
S_{\mathrm{chem}}^\Omega = - \sum_A c_A^\Omega \ln c_A^\Omega,
\end{equation}
which reaches its maximum for equiatomic mixtures and decreases upon segregation.

Finally, element-pair-resolved bond statistics were computed for each unordered chemical pair $(A,B)$ within region $\Omega$. Let $\mathcal{E}_{AB}^\Omega$ be the set of edges connecting atoms of species $A$ and $B$. The number of such bonds, their mean length, and their standard deviation were computed as
\begin{equation}
\begin{split}
&N_{AB}^\Omega = |\mathcal{E}_{AB}^\Omega|,\\
\qquad
&\langle d \rangle_{AB}^\Omega = \frac{1}{N_{AB}^\Omega} \sum_{(i,j) \in \mathcal{E}_{AB}^\Omega} d_{ij}, \qquad \\
&\sigma_{AB}^\Omega = \sqrt{\frac{1}{N_{AB}^\Omega} \sum_{(i,j) \in \mathcal{E}_{AB}^\Omega} (d_{ij} - \langle d \rangle_{AB}^\Omega)^2 }.
\end{split}
\end{equation}

All descriptors were evaluated independently for each structure, yielding a comprehensive, physically interpretable feature representation suitable for statistical analysis and machine-learning modeling of chemically complex materials.


The fragmented, layer-resolved descriptor is defined as
\begin{equation}
\mathbf{D} = \bigcup_{L=0}^{L_{\max}} \mathbf{D}^{(L)},
\end{equation}
where $\mathbf{D}^{(L)}$ denotes the set of features associated with topological layer $L$, ranging from the surface ($L=0$) to the deepest core layers.

To probe the relative importance of different regions of the nanoparticle, the descriptor can be written in weighted form as
\begin{equation}
\mathbf{D}_w = \sum_{L=0}^{L_{\max}} w_L \, \mathbf{D}^{(L)},
\end{equation}
where $w_L$ is a user-defined weight controlling the contribution of layer $L$ to the final representation.

Each layer-specific descriptor $\mathbf{D}^{(L)}$ contains averaged geometric, topological, and chemical features,
\begin{equation}
\mathbf{D}^{(L)} = \left\{ 
\langle z \rangle^{(L)}, 
\langle r \rangle^{(L)}, 
\mathbf{c}^{(L)}, 
\mathbf{p}^{(L)}, 
\mathbf{b}^{(L)}
\right\},
\end{equation}
where $\langle z \rangle^{(L)}$ is the mean coordination number, $\langle r \rangle^{(L)}$ denotes bond-length statistics, $\mathbf{c}^{(L)}$ the local chemical composition, $\mathbf{p}^{(L)}$ the nearest-neighbor pair probabilities, and $\mathbf{b}^{(L)}$ the corresponding bond-length distributions.

\subsection{Computational Details}

First-principles calculations were carried out within the framework of Density Functional Theory (DFT) as implemented in the \textsc{SIESTA} code. The metallic nanoparticle was modeled in a cubic supercell with dimensions of $35 \times 35 \times 35$~\AA, sufficiently large to prevent interactions between periodic images. Brillouin zone sampling was performed using a $\Gamma$ point  symmetry. 

Electron--ion interactions were described using norm-conserving pseudopotentials, and the electronic wave functions were expanded in a localized atomic orbital basis set of double-zeta with polarization (DZP). Exchange--correlation effects were treated within the generalized gradient approximation (GGA) using the Perdew--Burke--Ernzerhof (PBE) functional. A real-space mesh cutoff of 250~Ry was employed.

The Kohn--Sham equations were solved by direct diagonalization. Self-consistency was achieved using density matrix mixing with a weight of 0.035 and a convergence tolerance of $10^{-4}$. Spin polarization was not considered.

Structural optimization was performed using the conjugate gradient method while keeping the supercell volume fixed. Atomic positions were relaxed until the maximum force on each atom was below 0.02~eV/\AA, with a maximum atomic displacement of 0.1~\AA\ per relaxation step.


\subsection{Machine-learning models, ranking strategy, and interpretability}

To establish a robust and data-efficient learning framework for ranking quasi-crystal nanoparticle configurations, we explored a hierarchy of supervised machine-learning models with increasing expressive power. As linear baselines, we employed Ridge regression~\cite{Hilt1977}, which provides a regularized least-squares reference and serves as a lower-bound model for performance. In addition, single decision trees were considered to capture nonlinear dependencies between structural descriptors and total energy for Density Functional Theory (DFT).

Starting from a common initial atomic arrangement, a total of 1000 nanoparticle configurations were fully relaxed using DFT. All nanoparticles share the same global chemical composition, Al$_{70}$Co$_{10}$Fe$_5$Ni$_{10}$Cu$_5$, corresponding to a decagonal quasicrystalline alloy motif, while differing in their internal atomic arrangements and local chemical environments. This strategy enables systematic sampling of a rugged configurational energy landscape at fixed stoichiometry, isolating the effect of atomic ordering on nanoparticle stability. The chosen composition is motivated by a series of experimental studies reporting promising performance of Al–Co–Fe–Ni–Cu alloys in applications such as high-sensitivity chemical sensing and catalytic decomposition of toxic molecules\cite{Mishra2023,Mishra2023b}. By focusing on a chemically relevant and experimentally investigated composition, the present dataset provides a realistic testbed for data-driven stability ranking and accelerated exploration of complex nanoparticle configurations.

While the single decision trees models offer interpretability and low computational cost, their predictive capacity is limited in the presence of highly heterogeneous and correlated descriptors, as is the case for chemically complex nanoparticles.

To overcome these limitations, we adopted gradient-boosted decision trees implemented in the XGBoost~\cite{Chen2016} framework as our primary model. XGBoost is particularly well suited for medium-sized datasets and structured descriptors, offering strong nonlinear modeling capabilities, built-in regularization, and robustness against overfitting. Model hyperparameters—including tree depth, learning rate, number of estimators, subsampling ratios, and regularization strengths—were optimized using Bayesian optimization with the Optuna~\cite{10.1145/3292500.3330701} library. This automated hyperparameter search ensures fair and reproducible model comparison while systematically identifying near-optimal configurations in a high-dimensional parameter space. 

Rather than focusing exclusively on absolute energy prediction, we formulated the learning task as a ranking problem, which is more aligned with high-throughput screening and accelerated discovery workflows. Model performance was therefore primarily evaluated using rank-based metrics, including the Spearman rank correlation coefficient~\cite{Spearman1904}, which probes global monotonic consistency between predicted and reference energies. In addition, we employed top-$k$ screening recall and regret metrics~\cite{Liu2009,10.5555/3104322.3104451} to quantify the model’s ability to identify the most stable nanoparticle configurations within a limited computational budget. 

Once a reliable ranking model is established, the ability to quantify how confidently low-energy structures can be identified from a pool of candidates becomes central. In this work, we exploit a large reference dataset of density-functional-theory–relaxed nanoparticle configurations, which is partitioned into disjoint subsets used for model training and for independent ranking tests that emulate a realistic screening scenario. Having demonstrated that the model can robustly identify the most stable structures within a limited screening budget, we extend this capability toward an active learning workflow. In the publicly released descriptor code, we provide routines to generate new candidate configurations through small random atomic displacements and controlled modifications of the local chemical arrangement, thereby exploring the neighborhood of known low-energy structures. These newly generated candidates are subsequently ranked by the trained model, and the most promising ones can be selected for additional first-principles validation. By iteratively augmenting the training set with these targeted evaluations, the model can be progressively refined, enabling efficient and guided exploration of configurational space without the need for exhaustive sampling.

To interpret the trained XGBoost models and identify the most influential structural features, we performed post hoc explainability analysis using SHAP (SHapley Additive exPlanations)~\cite{10.5555/3295222.3295230}. This analysis provides physically meaningful insights into how different descriptor components,such as coordination, bonding statistics, and chemical ordering in different layers—contribute to the learned ranking, thereby linking model predictions back to underlying materials physics.

\section{Results}
Before addressing any learning or ranking results, we first analyze the structural descriptor itself. The proposed representation is based on a chemical graph constructed from interatomic neighbor relations, which requires only a single physically motivated parameter: the cutoff radius used to define bonds. In this work, a cutoff multiplier of 1.2 relative to the natural atomic radii was adopted, as this value provides a stable and chemically meaningful connectivity. Tests performed with larger cutoff values, up to approximately 1.4, resulted in only marginal changes in the descriptor statistics, indicating that the representation is not overly sensitive to this parameter. However, beyond this range, spurious and nonphysical connections may appear, especially in low-coordination surface regions. Additionally, the local environment was partitioned into six topological layers from the surface toward the interior, which was found to be sufficient to capture structural heterogeneity without introducing unnecessary descriptor redundancy.

Additional analyses were performed to characterize the statistical structure
and physical consistency of the layer-resolved descriptor dataset.
Supplementary Figures S2, S3 and S4, report quality-control metrics, principal component
analysis of the descriptor space, correlation maps among physically interpretable
features, and quantitative comparisons between surface and bulk environments.
In particular, layer-resolved surface–bulk contrasts in atomic fractions and
chemical entropy highlight systematic segregation and disorder trends across the
nanoparticle ensemble. These analyses confirm that the proposed descriptor
captures meaningful structural and chemical heterogeneity beyond simple global
averages and provides a physically grounded basis for data-efficient learning.

\begin{figure}[b!]
\centering
\includegraphics[width=0.5\textwidth]{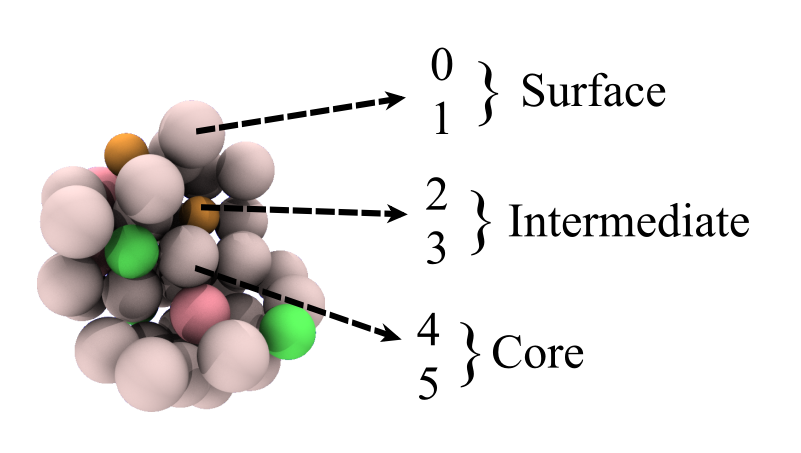}
\includegraphics[width=0.8\textwidth]{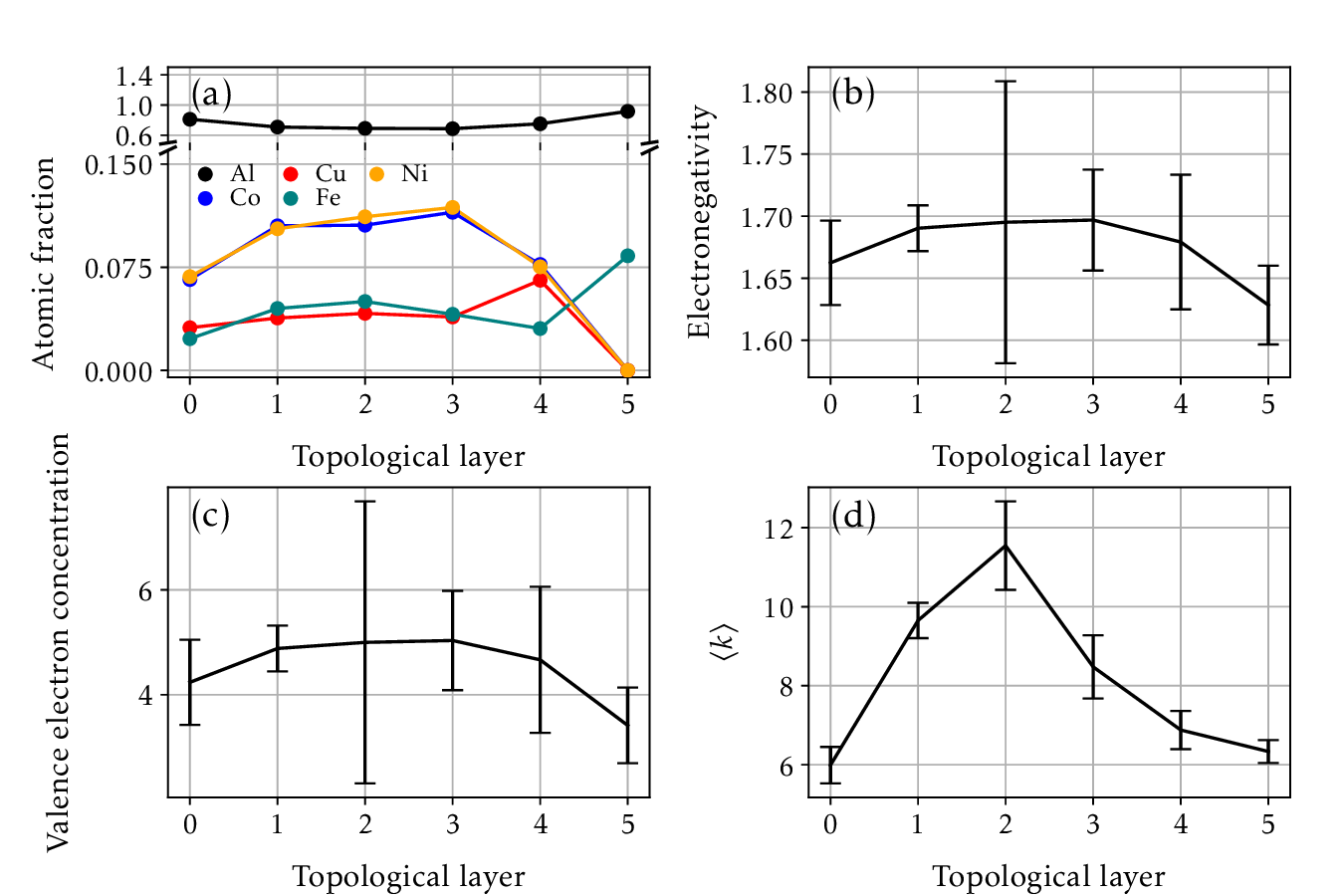}
\caption{Layer-resolved structural, chemical, and electronic descriptors for Al$_{70}$Co$_{10}$Fe$_5$Ni$_{10}$Cu$_5$ decagonal quasicrystalline alloy nanoparticles. 
Nanoparticles are partitioned into six topological layers defined by graph-based coordination analysis. 
Shown are (a) mean atomic fractions, (b) mean Pauling electronegativity, (c) mean valence electron concentration, and (d) mean coordination number as a function of topological layer index.}
\label{fig:mullayer}
\end{figure}

Figure~\ref{fig:mullayer} presents the layer-resolved decomposition of the proposed descriptor for Al$_{70}$Co$_{10}$Fe$_5$Ni$_{10}$Cu$_5$ decagonal quasicrystalline alloy nanoparticles. 

The Al$_{70}$Co$_{10}$Fe$_5$Ni$_{10}$Cu$_5$ nanoparticle composition investigated in this work was selected due to its relevance in recent experimental studies \cite{Mishra2023,Mishra2023b}, where alloys in this compositional space have demonstrated highly promising performance in advanced functional applications~\cite{Mandal2024,BeyramaliKivy2017,Aliyu2019}. In particular, multicomponent nanoparticles with similar elemental ratios have been reported to exhibit enhanced surface activity, chemical robustness, and tunable electronic properties, making them attractive platforms for high-sensitivity sensing~\cite{Chakraborty2025,Chakraborty2025a} and for catalytic processes involving the activation and decomposition of chemically hazardous species~\cite{Mishra2022,Choi2024,Manzoor2026}. These experimental indications motivate a detailed, atomistic-level investigation of the stability landscape of such nanoparticles, as understanding the interplay between atomic arrangement, chemical heterogeneity, and energetic stability is a prerequisite for rational design and optimization of their functional performance.

Each of 1000 nanoparticle is partitioned into six topological layers defined through a graph-based coordination analysis, enabling a physically meaningful separation of surface, subsurface, and core-like atomic environments without invoking explicit geometric or radial cutoffs. The mean number atoms by layer in shown in figure S1 in supplementary materials. This representation provides a compact yet expressive view of how structural, chemical, and electronic properties evolve across the nanoparticle.

Despite the fixed global stoichiometry, the layer-resolved mean atomic fractions reveal pronounced chemical inhomogeneity across the nanoparticle. The outermost layer is strongly enriched in Al, while transition-metal species are depleted relative to their bulk-averaged concentrations. This trend gradually weakens toward the interior, with deeper layers exhibiting a partial recovery of transition-metal content, as shown in figure \ref{fig:mullayer}-a). Such behavior is consistent with well-established segregation tendencies in multicomponent alloys, where elements with lower surface energy and larger atomic radius preferentially occupy undercoordinated surface sites. Importantly, the emergence of this segregation profile from a purely topological definition of layers demonstrates that coordination connectivity alone is sufficient to capture the dominant surface–core chemical gradients in quasicrystalline nanoparticles.

In figure \ref{fig:mullayer}-b) we shown the evolution of the mean Pauling electronegativity across layers mirrors this compositional redistribution. The surface layer exhibits a reduced average electronegativity, reflecting its Al-rich character, while subsurface layers display slightly higher and more uniform values. Toward the deepest layers, the electronegativity decreases again, indicating that the chemical environment of the nanoparticle core remains distinct from a homogeneous bulk average. The substantial variance observed in intermediate layers highlights the intrinsic configurational complexity of decagonal quasicrystalline alloys, where multiple chemically distinct local motifs coexist even at similar coordination depths.

A complementary picture emerges from the layer-dependent valence electron concentration, figure \ref{fig:mullayer}-c). The surface is characterized by a reduced VEC, consistent with Al enrichment and low coordination, whereas subsurface layers exhibit an increase toward transition-metal-dominated environments. Deeper layers show a gradual reduction in VEC, suggesting that electronic structure in the nanoparticle core reflects a nontrivial balance between composition, coordination, and finite-size effects. This non-monotonic behavior underscores that quasicrystalline nanoparticles cannot be accurately described as uniformly bulk-like systems, even in their interior regions.

Because electronegativity and valence electron concentration are composition-weighted averages, layers with lower Al content are more sensitive to fluctuations in the relative proportions of transition metals, as seen in topological layer 2, leading to larger statistical dispersion, whereas in Al-rich layers the dominant contribution of Al dampens these variations and reduces the error bars.

The structural foundation underlying these chemical and electronic trends is captured by the mean coordination number profile is shown in figure \ref{fig:mullayer}-d). Coordination increases sharply from the surface toward the interior, reaching a maximum in intermediate layers before decreasing slightly in the deepest layers considered. This behavior reflects the finite size of the nanoparticles and the absence of a fully bulk-like coordination environment. Crucially, it validates the topological layering scheme: each layer corresponds to a distinct and statistically meaningful coordination regime rather than an arbitrary geometric shell.

Before proceeding to the learning and ranking analysis, it is instructive to examine the intrinsic energetic dispersion of the dataset generated from chemically randomized 55-atom dodecahedral nanoparticles. Although all configurations share the same global stoichiometry and initial morphology, full DFT relaxation reveals a broad energetic landscape. The total energy distribution spans approximately 55–57~eV across the 1000 configurations, corresponding to roughly 1.0~eV per atom. This substantial energetic spread indicates that chemical ordering and layer-dependent segregation patterns generate distinctly different stability basins rather than small perturbations around a single minimum. In other words, even under fixed composition and morphology, the configurational landscape is highly non-trivial, providing a physically meaningful and information-rich foundation for data-efficient ranking. Having established that the dataset spans a broad and rugged energetic regime, we now analyze the structure of the descriptor space through principal component analysis.

\begin{figure}
\centering
\includegraphics[width=0.6\textwidth]{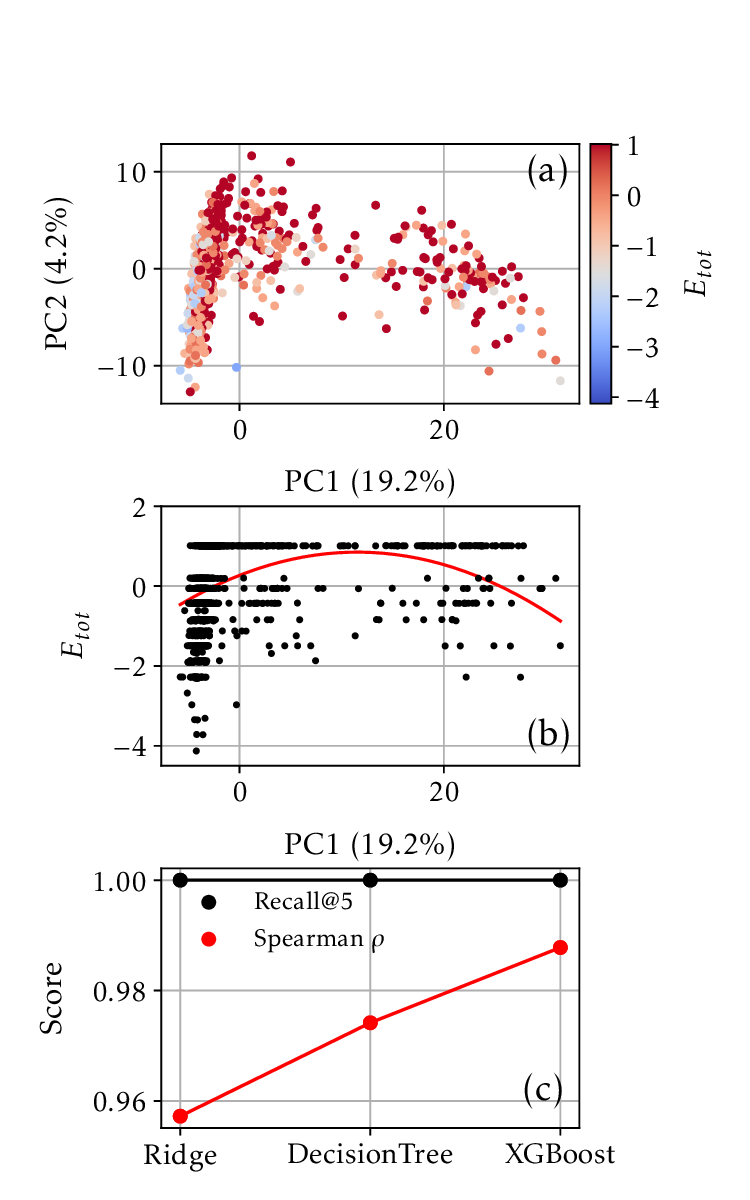}
\caption{Principal component analysis and model comparison for the ranking of Al$_{70}$Co$_{10}$Fe$_5$Ni$_{10}$Cu$_5$ decagonal quasicrystalline alloy nanoparticles.
(a) Projection of the descriptor space onto the first two principal components, colored by the total energy $E_{\mathrm{tot}}$.
(b) Total energy as a function of the first principal component, illustrating the non-linear relationship between dominant descriptor variance and energetic stability.
(c) Comparison of baseline regression models and Optuna-optimized XGBoost in terms of ranking performance, quantified by the Spearman rank correlation and Recall@5.}
\label{fig:pca}
\end{figure}

Figure~\ref{fig:pca} examines the structure of the descriptor space and its implications for model selection in the ranking of Al$_{70}$Co$_{10}$Fe$_5$Ni$_{10}$Cu$_5$ decagonal quasicrystalline alloy nanoparticles. A principal component analysis is first employed as a diagnostic tool to probe the intrinsic organization of the descriptor space prior to model training. The projection onto the first two principal components in figure \ref{fig:pca}-a) reveals a structured but strongly overlapping distribution of configurations when colored by the total energy. In figure \ref{fig:pca}-b), although PC1 captures a non-negligible fraction of the total variance, low- and high-energy structures remain significantly intermixed, indicating that the relationship between the descriptor and the total energy is not governed by a simple linear trend in the dominant variance directions. This observation provides a clear physical justification for moving beyond linear regression models.

In figure \ref{fig:pca}-c), we show the comparison of baseline regression models further clarifies this point. Linear ridge regression already achieves a relatively high Spearman rank correlation, indicating that part of the energetic ordering can be captured by linear combinations of the descriptors. A single decision tree improves upon this performance by introducing non-linear thresholds, reflecting the inherently non-linear coupling between coordination, chemical composition, and electronic descriptors in quasicrystalline nanoparticles. However, both models remain limited in their expressivity, as they rely either on global linear trends or on a small number of hierarchical splits.

The XGBoost model optimized via Bayesian hyperparameter tuning with Optuna yields the highest overall rank correlation, demonstrating its superior ability to learn complex, multi-scale interactions among the descriptors. While the Recall@5 metric is slightly lower for XGBoost in this particular train--test split, the regret remains strictly zero across all models and screening budgets considered. This indicates that, despite minor differences in the exact composition of the top-ranked subset, all approaches consistently identify at least one structure belonging to the true lowest-energy basin. Importantly, the higher Spearman correlation achieved by XGBoost implies a more reliable global ordering of configurations, which is critical for stability ranking and for iterative workflows such as active learning, where errors propagate across successive selection steps.

As discussed previously, the atomic-scale description of nanoparticles can be naturally decomposed into topological layers defined by the distance from the surface. This representation provides a physically transparent framework to investigate how structural, geometric, and chemical descriptors vary across different regions of the nanoparticle and how each region contributes to the total energy. Such a layered view is particularly appealing for multicomponent alloys, where coordination, bonding, and chemical short-range order may differ substantially between surface, subsurface, intermediate, and core-like environments.

As detailed in the Methodology section and illustrated in Fig.~\ref{fig:lc_multitest}, the proposed fragmented descriptor enables an explicit partitioning of each nanoparticle into multiple concentric layers, providing a physically motivated representation of its internal structure. In the present work, we consider a decomposition into six topological layers. Layer $L0$ corresponds to the outermost surface, directly exposed to the environment, while $L1$ represents the subsurface layer that is chemically and structurally coupled to the surface. Layers $L2$ and $L3$ define intermediate regions that progressively transition from surface-like to bulk-like coordination environments. Finally, layers $L4$ and $L5$ correspond to the most internal atoms, collectively forming the nanoparticle core. This layered description allows us to systematically probe how geometric, chemical, and bonding characteristics evolve from the surface toward the core, and to assess the relative contribution of each region to the overall energetic stability of the nanoparticle.

\begin{figure}
\centering
\includegraphics[width=0.5\linewidth]{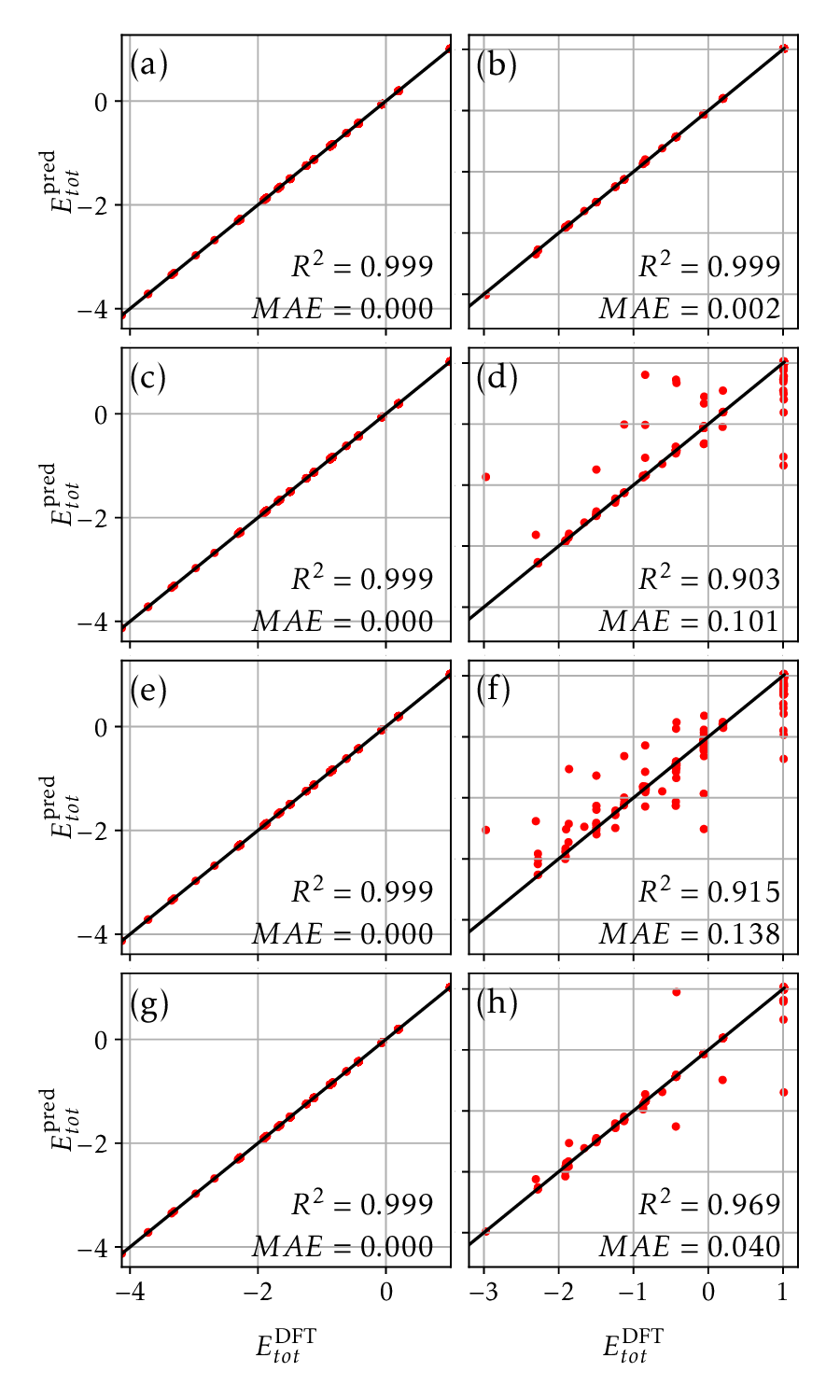}
\caption{Training (left) and test (right) performance of the Optuna-optimized XGBoost model under different layer-weighting schemes.
(a,b) Uniform weighting of all topological layers ($w_L = 1.0$).
(c,d) Surface-emphasized embedding with $w_{L0}=w_{L1}=1.0$ and reduced weights for deeper layers.
(e,f) Intermediate-layer–emphasized embedding with $w_{L2}=w_{L3}=1.0$.
(g,h) Core-emphasized embedding with $w_{L4}=w_{L5}=1.0$.
Each panel reports predicted versus DFT total energies, illustrating how the relative emphasis on surface, intermediate, or core environments affects model generalization.}
\label{fig:xg}
\end{figure}

Figure~\ref{fig:xg} summarizes the predictive performance of the XGBoost model trained on the proposed descriptor under different layer-weighting schemes. Panels \ref{fig:xg}-a) and \ref{fig:xg}-b) correspond to the reference case in which all layers are assigned equal weight ($w_L = 1.0$ for all $L$).

In this configuration, the model achieves excellent agreement with DFT total energies, with near-perfect performance on the training set and very high accuracy on the held-out test set. This result is not unexpected given the size and quality of the available dataset, which consists of approximately $10^3$ fully relaxed DFT calculations spanning a dense and well-sampled configurational space. In such a regime, gradient-boosted tree models are able to effectively learn complex, nonlinear relationships between structural descriptors and total energy, as already established by the comparative analysis presented in the previous figure, where XGBoost consistently outperformed linear and single-tree models.

The uniform-weight case serves as a critical baseline: it demonstrates that the descriptor, when all layers are treated on equal footing, contains sufficient information to accurately reconstruct the energetic landscape of the system. Importantly, this also validates the layered descriptor itself, showing that the partition into topological shells does not introduce spurious noise or degrade predictive power relative to more monolithic representations.

Panels \ref{fig:xg}-c) and \ref{fig:xg}-d) explore a surface-emphasized representation, in which the outermost layers ($L_0$ and $L_1$) are assigned full weight, while deeper layers are down-weighted. In this scenario, the training performance remains essentially ideal, reflecting the high expressive capacity of the model. However, a noticeable degradation is observed on the test set, with an increase in the mean absolute error and a reduction in $R^2$. This behavior indicates that although surface descriptors are strongly correlated with stability, they are not sufficient on their own to fully capture the energetic ordering across the entire dataset. In other words, while surface chemistry and coordination play a crucial role, especially in small nanoparticles, neglecting or suppressing information from the interior leads to a loss of generalization.

A similar trend is observed in panels \ref{fig:xg}-e) and \ref{fig:xg}-f), where the emphasis is shifted toward intermediate layers ($L_2$ and $L_3$). The resulting performance is comparable to, but slightly worse than, the surface-focused case on the test set. This suggests that the mid-shell region alone does not uniquely encode the energetic fingerprints required for robust prediction. Instead, it appears to act as a transition zone whose descriptors are informative only when combined with both surface and deeper-layer information.

Finally, panels \ref{fig:xg}-g) and \ref{fig:xg}-h) correspond to a core-emphasized representation, where the deepest layers ($L_4$ and $L_5$) are assigned full weight. Interestingly, this configuration recovers a significantly better test-set performance than the surface- or mid-layer–focused schemes, although it still does not fully match the uniform-weight baseline. This result highlights that even in nanoparticles of moderate size, core-like environments retain a strong energetic signature that correlates with overall stability. At the same time, the residual gap relative to the equal-weight case underscores that the total energy emerges from a collective interplay between surface, intermediate, and core regions, rather than being dominated by a single spatial zone.

Although the model exhibits clear indications of overfitting, evidenced by the nearly perfect fit on the training set ($R^2 \approx 1.000$ and $\mathrm{MAE} \approx 0$) and the consistent reduction in performance on the test set, this behavior is expected given the high flexibility of XGBoost and the nature of the problem, in which structures with the same chemical composition display relatively smooth and correlated energy variations. More importantly, the introduction of different nanoparticle partitioning schemes does not significantly alter the magnitude of the errors on the test set, which remain comparable across all cases. This indicates that the eventual overfitting affects the models in a systematic and homogeneous manner, without compromising the proposed comparative analysis. 

However, this behavior must be interpreted in light of both the intrinsic characteristics of the system and the controlled model configuration. First, all models were optimized using Optuna, resulting in highly consistent sets of hyperparameters — for instance, in the surface model ($\texttt{max\_depth} = 8$, $\texttt{learning\_rate} \approx 0.053$, $\texttt{n\_estimators} =1298$, $\texttt{subsample} \approx 0.97$, $\texttt{reg\_lambda} \approx 0.10$), ensuring that comparisons among partitioning schemes are performed under the same regime of complexity and regularization. Furthermore, the system under investigation is physically continuous: the structural configurations share the same chemical composition and differ only in atomic arrangement. Consequently, the model operates within a highly correlated feature space, which favors an extremely accurate fit on the training data without necessarily impairing generalization within the same structural distribution. 

The fact that test errors remain stable and comparable across different partitioning strategies indicates that the observed overfitting does not introduce differential bias among the models. Rather than representing detrimental overfitting, the results reflect the strong expressive capacity of XGBoost to capture the energy–structure relationship within this specific regime. Since the primary objective of this work is not to maximize absolute predictive accuracy, but to consistently evaluate the relative importance of each nanoparticle layer in the predictive performance, the obtained results are methodologically appropriate and sufficiently robust to support the proposed conclusions.

\begin{figure}
\centering
\includegraphics[width=0.5\textwidth]{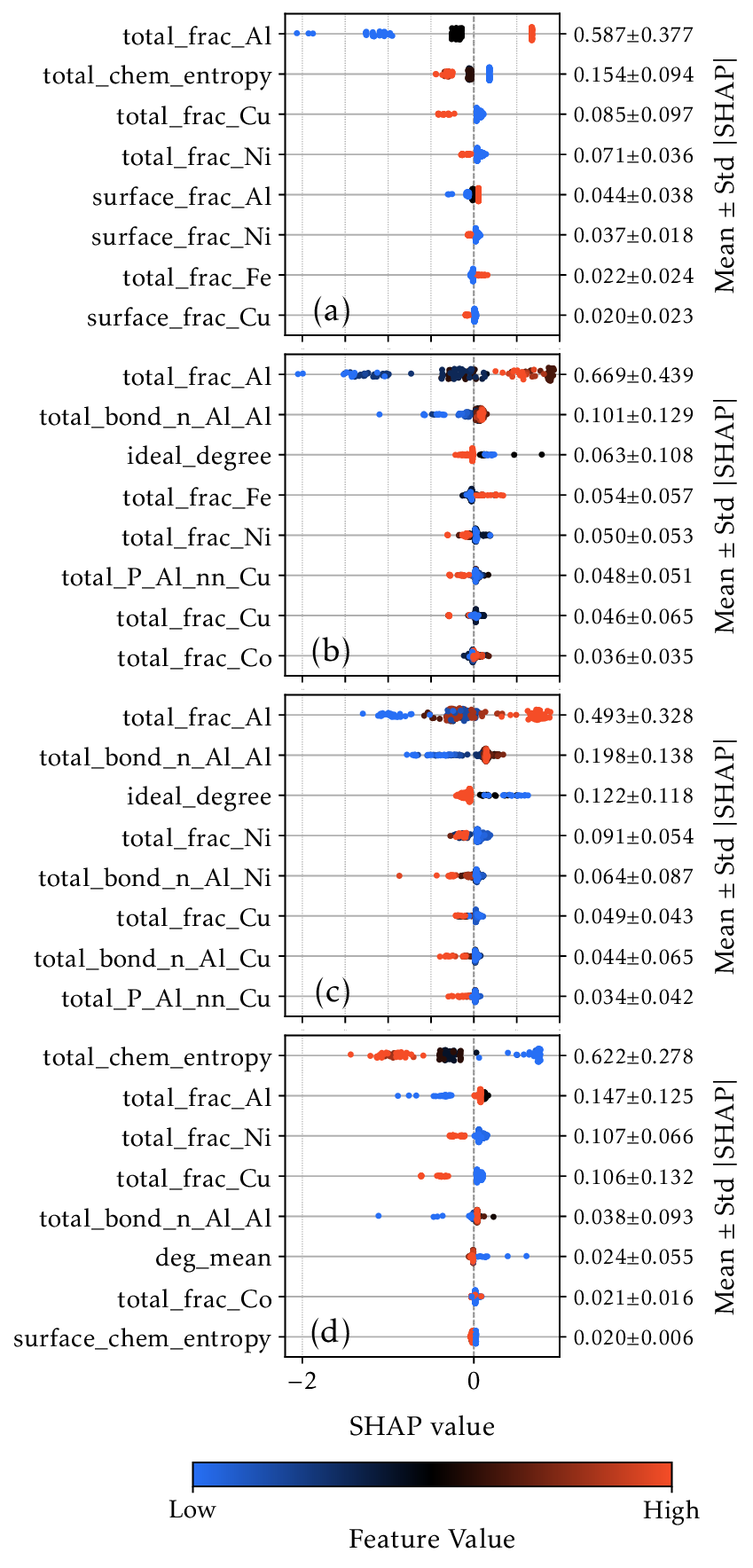}
\caption{SHAP (SHapley Additive exPlanations) analysis of the Optuna-optimized XGBoost model for different layer-weighting schemes, shown in the same top-to-bottom order as Fig.~\ref{fig:xg}.
(a) Uniform weighting of all topological layers.
(b) Surface-emphasized embedding.
(c) Intermediate-layer–emphasized embedding.
(d) Core-emphasized embedding.
Left panels show SHAP summary plots, reporting the distribution of feature contributions across the dataset, while right panels rank features by mean absolute SHAP value, highlighting the dominant structural and chemical descriptors controlling the predicted total energy.}
\label{fig:shap}
\end{figure}

Building upon the ranking performance established in Fig.~\ref{fig:xg}, we next assess the interpretability of the learned models through SHAP (SHapley Additive exPlanations) analysis, as summarized in Fig.~\ref{fig:shap} following the same top-to-bottom ordering. While the introduction of layer-dependent weighting naturally increases the flexibility of the model representation, the predictive performance remains stable across all embedding strategies. Minor variations between training and test metrics are observed when specific regions of the nanoparticle are emphasized, but these do not compromise the overall ranking capability of the model. Importantly, this controlled increase in model expressiveness enables a detailed, physically meaningful interpretation of how different structural and chemical features contribute to stability in distinct spatial regions. The combination of XGBoost with SHAP analysis allows us to systematically disentangle the roles of surface, intermediate, and core environments, revealing region-specific drivers of energetic stability without sacrificing predictive reliability. This balance between performance and interpretability is central to the proposed framework and highlights its suitability for exploratory and data-efficient studies of complex nanoparticles.

Figure~\ref{fig:shap} reports a SHAP (SHapley Additive exPlanations) analysis for the Optuna-optimized XGBoost model, shown in the same top-to-bottom ordering as Fig.~\ref{fig:xg}: uniform weighting (all layers), surface-emphasized, medium-layer–emphasized, and core-emphasized embeddings. SHAP provides a principled, model-agnostic attribution of how each input feature contributes to the prediction for each structure, based on Shapley values from cooperative game theory. In practice, the SHAP ``summary'' panels (left column) visualize, for each feature, the distribution of its contributions across the test set, while the bar plots (right column) rank features by mean absolute contribution, $\langle|\text{SHAP}|\rangle$, thus providing a compact measure of global importance. This interpretability layer is essential in our context: it allows us to connect the learned ranking/energy model back to physically meaningful descriptors (coordination, local chemistry, bonding statistics) and to assess which nanoparticle regions (surface, intermediate, core) control the energetic ordering.

In the uniform-weight baseline (panel \ref{fig:shap}-a), the dominant drivers are composition-like and disorder-like proxies, with $\texttt{total\_frac\_Al}$ and $\texttt{total\_chem\_entropy}$ emerging as the leading contributors, followed by $\texttt{total\_frac\_Cu}$ and $\texttt{total\_frac\_Ni}$, and smaller but non-negligible surface-fraction terms (e.g., $\texttt{surface\_frac\_Al}$, $\texttt{surface\_frac\_Ni}$ and $\texttt{surface\_frac\_Cu}$). 

Because the global stoichiometry of the decagonal quasicrystalline alloy is fixed across the dataset, these ``fraction'' features should be interpreted not as trivial composition labels, but as \emph{effective, environment-weighted occupancies} that encode segregation and redistribution between topological shells. In other words, even at fixed overall composition, the energetically relevant signal is how species populate different coordination environments and how chemical disorder manifests locally. The prominence of $\texttt{total\_chem\_entropy}$ indicates that the model systematically leverages descriptors capturing local mixing/heterogeneity---a natural energetic lever in complex alloys, where stability is shaped by a competition between preferred unlike/like neighborhoods and the frustration inherent to quasicrystalline order.

When the embedding is biased toward the surface (panel b), the importance ranking shifts toward features that explicitly encode surface-sensitive chemical and bonding motifs, while still retaining strong dependence on $\texttt{total\_frac\_Al}$ and introducing topological/bonding terms such as $\texttt{total\_bond\_n\_Al\_Al}$ and neighborhood statistics (e.g., $\texttt{total\_P\_Al\_nn\_Cu}$), along with $\texttt{ideal\_degree}$. 
This is physically consistent: emphasizing $L_0$--$L_1$ increases the relative weight of under-coordinated environments, where surface segregation, local short-range order (SRO), and specific bond populations can exert an amplified influence on the total energy. Importantly, this surface-focused SHAP portrait complements Fig.~\ref{fig:xg}-c/d): while the surface-weighted model can achieve strong apparent fit, its generalization is slightly more fragile, which is reflected by the emergence of sharper, more specific descriptors (e.g., individual bond-count terms) as key predictors. This is a typical signature of a model operating with a more restricted ``view'' of the system: it becomes more sensitive to particular motifs that correlate well within the training set, increasing the risk of mild overfitting relative to the all-layer baseline.


The medium-layer emphasis (panel \ref{fig:shap}-c) further reinforces this motif-level interpretability: bonding populations involving Al (e.g., $\texttt{total\_bond\_n\_Al\_Al}$, $\texttt{total\_bond\_n\_Al\_Cu}$, $\texttt{total\_bond\_n\_Al\_Ni}$) appear among the most influential variables, together with $\texttt{ideal\_degree}$ and accompanied by $\texttt{total\_frac\_Ni}$ and $\texttt{total\_frac\_Cu}$.
This behavior suggests that the intermediate shells are a region where the model benefits from descriptors that explicitly couple chemistry and topology: the mid-layers act as a bridge between surface-driven coordination deficits and core-like packing constraints. In this regime, the model highlights bond-population features as a compact summary of how the quasicrystalline alloy accommodates chemical frustration in partially coordinated environments. The fact that both $\texttt{ideal\_degree}$ and entropy-related terms, such as hints to the local mixing of atomic species, remain important indicates that the energetic ordering is not governed by a single local motif, but by an interplay between (i) how ``bulk-like'' the coordination network is and (ii) how chemical disorder is distributed across the nanoparticle.

Finally, in the core-emphasized embedding (panel \ref{fig:shap}-d), the SHAP ranking reveals a coherent picture:$\texttt{deg\_mean}$ becomes particularly informative, and surface-only descriptors such as $\texttt{surface\_chem\_entropy}$ and $\texttt{surface\_frac\_Ni}$ enter the list alongside $\texttt{total\_chem\_entropy}$ and the effective fractions $\texttt{total\_frac\_Al}$, $\texttt{total\_frac\_Ni}$, and $\texttt{total\_frac\_Cu}$.
At first glance, the presence of surface terms in a core-emphasized model may seem counterintuitive; however, it is a natural consequence of nanoparticle energetics being globally constrained: even if the descriptor is weighted toward deeper shells, the energy still reflects a coupled equilibrium between core packing/coordination and surface relaxation/segregation. In other words, the core cannot be interpreted in isolation because the nanoparticle must satisfy global stoichiometry and mechanical/chemical compatibility between regions. This is precisely where the layered embedding becomes valuable: it enables controlled \emph{what-if} experiments showing which descriptors remain important when a given region is preferentially emphasized.

A central advantage of the proposed framework is therefore not merely predictive accuracy, but \emph{diagnostic capability}. By fragmenting the descriptor into topological shells and re-embedding it via controlled layer weights, we can probe the relative energetic relevance of surface, intermediate, and core environments while keeping the feature dimensionality fixed (crucial for robust learning at $\sim 10^3$ DFT data points). The mild increase in overfitting tendency observed in the fragmented (region-emphasized) cases relative to the uniform baseline is expected: reweighting reduces the effective information content available to the model and amplifies region-specific correlations. Nevertheless, performance remains strong, and the interpretability gain is substantial: the SHAP analyses directly identify which chemical/topological motifs dominate in each regime, providing a physically grounded narrative of stability in quasicrystalline alloy nanoparticles.

More broadly, this layer-resolved interpretability is especially relevant for nanoscale alloy design, where systems frequently develop heterogeneous architectures such as \emph{core--shell} (or more generally compositionally graded) nanoparticles, in which an inner region (core) exhibits a different effective stoichiometry/ordering than the outer region (shell). In such cases, the ability to isolate and quantify region-specific energetic drivers is crucial for rational optimization. Here, even under fixed global stoichiometry, the SHAP results demonstrate that effective layer-resolved occupancies, coordination topology, and local chemical disorder jointly control $E_{\text{tot}}$; this supports the broader applicability of our method to nanoparticles where compositional partitioning and shell-specific motifs are an intentional design variable.

\section{Data-efficient discovery of stable alloy nanoparticles via ranking and active learning}

Building upon the layer-resolved descriptor analysis and the interpretability insights provided by the XGBoost–SHAP framework, we now turn to the central practical objective of this work: the efficient identification of low-energy nanoparticle configurations through a data-efficient ranking strategy and its natural extension to active learning.

We investigate a chemically and complex space of nanoparticles with following composition Al$_{70}$Co$_{10}$Fe$_5$Ni$_{10}$Cu$_5$, a total of 1000 distinct nanoparticle configurations were generated and fully relaxed using density functional theory (DFT) calculations performed with the \textsc{SIESTA} code~\cite{Soler2002}, as detailed in the Methodology section.

These structures differ in atomic arrangement and local chemical environments while preserving the same global stoichiometry, resulting in a highly rugged potential energy landscape. The total energy per atom, $E_{\text{tot}}$, is used as a stability metric, as lower energies correspond to thermodynamically more favorable nanoparticle configurations. Despite the fixed composition, the combinatorial number of possible atomic arrangements leads to an enormous configurational space, making exhaustive first-principles exploration computationally prohibitive.

Rather than aiming at direct energy prediction, we formulate the problem as a \emph{ranking task}, whose primary goal is to reliably identify low-energy candidates among a large pool of possible structures. This perspective naturally aligns with accelerated discovery strategies such as active learning, where models iteratively propose new nanoparticle configurations expected to exhibit enhanced stability. By ranking structures according to their predicted energetic favorability, the method enables efficient down-selection of promising candidates, dramatically reducing the number of expensive DFT calculations required. Crucially, this approach allows one to traverse the vast configurational space of nanoparticles in a targeted manner, focusing computational resources on regions most likely to yield stable structures.

Given the size of the available dataset (1000 fully relaxed nanoparticles), we deliberately avoid deep neural network architectures, which typically require substantially larger training sets to generalize reliably in high-dimensional descriptor spaces. Instead, we employ gradient-boosted decision trees using the XGBoost framework, combined with Bayesian hyperparameter optimization via Optuna. XGBoost offers several advantages in this regime: strong performance with limited data, robustness to heterogeneous and partially correlated descriptors, and inherent regularization that mitigates overfitting. These properties make it particularly well suited for learning structure--energy relationships in complex materials systems with constrained data availability. This choice enables a systematic investigation of how ranking performance evolves as a function of training set size, providing critical insights into the data efficiency and practical applicability of the proposed framework. The structural descriptors were constructed by partitioning each nanoparticle into only two regions, bulk and surface, since additional subsurface layering was found not to introduce significant new information while substantially increasing descriptor complexity.

\begin{figure}
\centering
\includegraphics[width=0.5\linewidth]{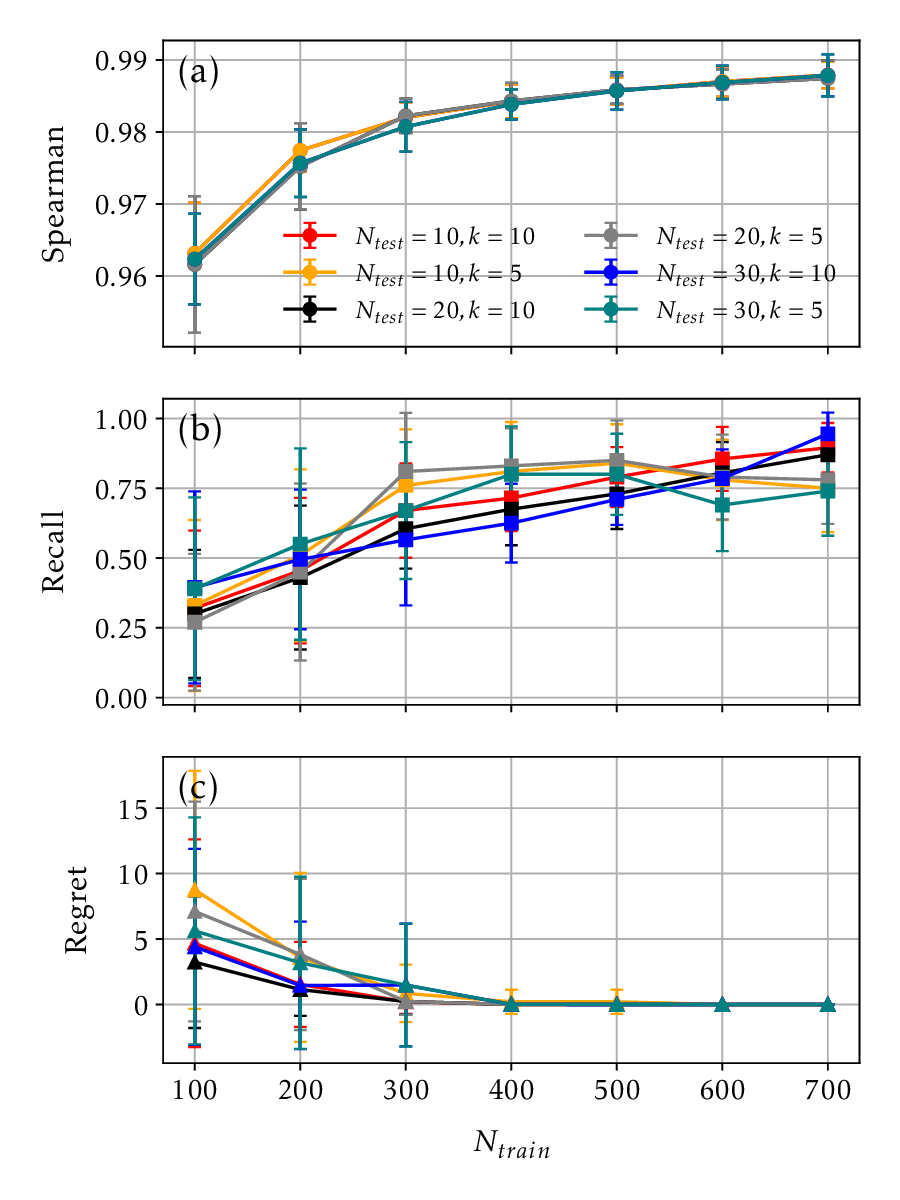}
\caption{Learning curves for the ranking of AlFeCoNiCu nanoparticle configurations as a function of the training-set size $N_{\mathrm{train}}$. 
Results are shown for three test-set sizes ($N_{\mathrm{test}} = 10, 20, 30$) and two screening budgets ($k = 5, 10$). 
(a) Spearman rank correlation $\rho$ between predicted and DFT reference energies, probing global rank consistency. 
(b) Top-$k$ screening recall, quantifying the fraction of true low-energy structures recovered within the model-selected subset. 
(c) Top-$k$ regret, defined as the energy difference between the lowest-energy structure identified within the selected top-$k$ and the true global minimum. 
Error bars indicate the standard deviation over multiple random train--test splits and model initializations.}
\label{fig:lc_multitest}
\end{figure}

Figure~\ref{fig:lc_multitest} summarizes the ranking performance as a function of training-set size ($N_{\text{train}}$) for three fixed test-set sizes ($N_{\text{test}}=10,20,30$) and two screening budgets ($k=5,10$). We report (a) the Spearman rank correlation $\rho$ between predicted and reference energies, (b) the top-$k$ screening recall, and (c) the top-$k$ regret (energy loss with respect to the true optimum), with error bars reflecting variability across repeated random splits/seeds.

Across all conditions, Spearman $\rho$ increases monotonically with $N_{\text{train}}$ and quickly approaches a high-correlation regime (near-saturation by $N_{\text{train}}\approx 300$--$500$). This behavior indicates that the model learns a stable global ordering of candidate structures once a modest number of labeled (DFT) points are available. Importantly, the $\rho$ curves for $N_{\text{test}}=10,20,30$ are nearly indistinguishable within uncertainty at fixed $N_{\text{train}}$, implying that the \emph{expected} correlation estimate is robust to the specific test-set size in this range. The main role of $N_{\text{test}}$ is therefore statistical: larger $N_{\text{test}}$ provides slightly tighter uncertainty bounds (especially at small $N_{\text{train}}$) because the correlation is estimated from more held-out samples. Finally, changing the screening budget from $k=5$ to $k=10$ does not materially alter $\rho$, which is expected because $\rho$ probes the \emph{global} rank consistency rather than the extreme top of the list.

While Spearman $\rho$ captures global monotonicity, the screening recall directly evaluates the use-case of interest: identifying a small number of best candidates for follow-up calculations. The recall curves show strong gains with training size, but the sensitivity to $k$ is notably larger than for $\rho$. At low-to-moderate $N_{\text{train}}$ (e.g., $100$--$300$), recall for $k=5$ is often higher than for $k=10$ in our setting. This result is consistent with the fact that increasing $k$ makes the evaluation stricter when the target set is defined as the \emph{true} top-$k$: the model must correctly place more near-optimal structures within the selected subset, which is challenging when the training signal is still limited and when many candidates have closely spaced energies. In contrast, once $N_{\text{train}}$ reaches the regime where the ranking stabilizes ($\gtrsim 400$), recall becomes high for both budgets and differences between $k=5$ and $k=10$ narrow, indicating that the model not only identifies the single best candidate but also captures a broader near-optimal set.

For $N_{\text{test}}=10$, recall estimates exhibit visibly larger dispersion, especially at small $N_{\text{train}}$, because each held-out structure represents a large fraction of the test set. Increasing $N_{\text{test}}$ to $20$ and $30$ reduces the discretization/noise in recall (and yields more stable means), making performance comparisons between $k$ values more reliable. This observation motivates reporting at least $N_{\text{test}}\ge 20$ for publication-quality estimates, while keeping $N_{\text{test}}$ fixed across learning-curve points to ensure fair comparisons.

Regret provides a stringent, decision-centric metric: it measures the energy gap between the best structure found within the model-selected top-$k$ and the true global minimum. At small $N_{\text{train}}$, regret is non-zero and exhibits large variance across splits, reflecting occasional failures where the model misses the optimal basin. However, regret decreases sharply with $N_{\text{train}}$ and becomes essentially zero by $N_{\text{train}}\approx 300$--$400$ for all $N_{\text{test}}$ and both $k$ values. This is a key practical outcome: in the data-rich regime, the pipeline consistently recovers the true optimum \emph{within the top-$k$ set}, meaning that follow-up computations restricted to only a handful of candidates would still identify the best structure.

In high-throughput materials discovery, $k$ corresponds to the \emph{experimental/DFT screening budget}: the number of candidates one can afford to validate with expensive calculations. Our results show that $k=5$ offers an excellent balance: it yields strong recall and rapidly vanishing regret at moderate $N_{\text{train}}$, while maintaining a stringent and practically meaningful selection size. Increasing to $k=10$ does not improve the global ranking correlation and can reduce recall at low-to-moderate training sizes because the method must correctly identify a larger near-optimal set, which is intrinsically harder when many structures are energetically similar. Therefore, $k=5$ is not an arbitrary choice; it is empirically supported by superior screening performance in the regime where screening efficiency matters most, and it matches realistic computational budgets.

Beyond accuracy metrics, an essential question for accelerated materials discovery is \emph{how much reference data is actually needed to reach actionable performance}. The learning curves in Fig.~\ref{fig:lc_multitest} demonstrate that the proposed descriptor–learning framework achieves near-optimal behavior with a remarkably small number of training structures. Already at $N_{\text{train}}\approx 200$--$300$, the model reaches a regime of high rank correlation ($\rho \gtrsim 0.97$), near-unity Recall@$5$, and vanishing regret, indicating reliable identification of the true minimum using only a handful of screened candidates. Increasing the training set beyond this point yields only marginal gains, signaling an early saturation of performance. This data efficiency is particularly significant in the context of first-principles screening, where each additional labeled structure incurs a substantial computational cost. As a result, the method enables a realistic workflow in which a few hundred reference calculations suffice to guide the exploration of thousands of candidate nanostructures with high confidence. Taken together, these results position the present approach as a robust, scalable, and cost-effective solution for high-throughput ranking and down-selection in complex multicomponent materials spaces.

In addition to ranking existing configurations, the present framework naturally enables an active learning workflow, in which the trained XGBoost model is iteratively used to propose new candidate nanoparticle structures by perturbing atomic configurations or compositions and prioritizing those predicted to minimize a target property, such as the total energy per atom. Importantly, the implementation is fully general: the same pipeline can be applied to any user-defined descriptor and any scalar target, independent of the underlying electronic-structure method. To promote reproducibility and facilitate adoption by the community, we make the complete codebase publicly available at (https://github.com/tromer-unb/layer-resolved-ml-nanoparticles/tree/main), including routines for descriptor generation, model training, ranking, and automated proposal of new candidate structures.

\section{Summary and Conclusions}

In this work, we presented a data-efficient and physically interpretable machine-learning framework for exploring the configurational stability of chemically complex nanoparticles. By explicitly decomposing nanoparticles into topological layers defined by their connectivity to the surface, we introduced a fragmented descriptor that retains essential spatial resolution while preserving a compact and fixed feature dimensionality. This representation directly addresses a central bottleneck in nanoparticle modeling: capturing heterogeneous surface and interior environments without incurring the high computational cost and data requirements of fully local atom-centered approaches.  

Using gradient-boosted decision tree models and formulating the learning task as a ranking problem, we demonstrated that accurate identification of the most stable nanoparticle configurations can be achieved with only a few hundred first-principles reference calculations. Ranking-based metrics show rapid saturation of performance, with high rank correlation, strong top-$k$ recall, and vanishing regret at moderate training-set sizes. These results highlight that reliable down-selection of low-energy candidates does not require exhaustive sampling of the configurational space, making the proposed approach well suited for realistic high-throughput screening scenarios.  

Beyond predictive performance, the layer-resolved nature of the descriptor provides direct physical insight into how different regions of the nanoparticle contribute to energetic stability. Through controlled layer weighting and SHAP-based interpretability analysis, we showed that surface segregation, coordination topology, chemical disorder, and bonding motifs play distinct and coupled roles across surface, intermediate, and core environments. This ability to disentangle region-specific energetic drivers represents a key advantage over monolithic descriptor frameworks and enables mechanistic interpretation of stability trends in complex multicomponent nanoparticles.  

From a broader perspective, the proposed framework naturally supports active learning strategies for nanoscale materials discovery. Once trained, the ranking model can efficiently guide additional first-principles calculations toward the most promising regions of configurational space, progressively refining the stability landscape at minimal computational cost. Owing to its general formulation, the approach is not restricted to a specific alloy composition or nanoparticle morphology and can be readily extended to other multicomponent nanostructures, including systems with compositional gradients, core--shell architectures, or chemically distinct surface terminations.  

Overall, this work establishes a scalable, interpretable, and computationally efficient pathway for accelerated discovery and understanding of stable nanostructures, bridging the gap between first-principles accuracy and data-driven exploration in complex nanoscale materials systems.

\section{Acknowledgements}

This work was financed in part by the Coordination for the Improvement of Higher Education Personnel (CAPES), the Brazilian National Council for Scientific and Technological Development (CNPq), the Araucaria Foundation, and FAPESP. CFW acknowledges CNPq for a Research Productivity Fellowship (grant no. 310603/2025-0), and RMT acknowledges CNPq for a Research Productivity Fellowship (grant no. 307371/2025-5). The authors thank the Coaraci Supercomputer for computer time (FAPESP grant no. 2019/17874-0) and the Center for Computing in Engineering and Sciences at Unicamp (FAPESP grant no. 2013/08293-7). We further acknowledge the National Laboratory for Scientific Computing (LNCC/MCTI, Brazil) for providing HPC resources of the SDumont supercomputer, which contributed to the research results presented in this work.

\bibliography{bibliography}



\end{document}



\section{Supplementary figures}

\begin{figure}[b!]
\centering
\includegraphics[width=0.75\linewidth]{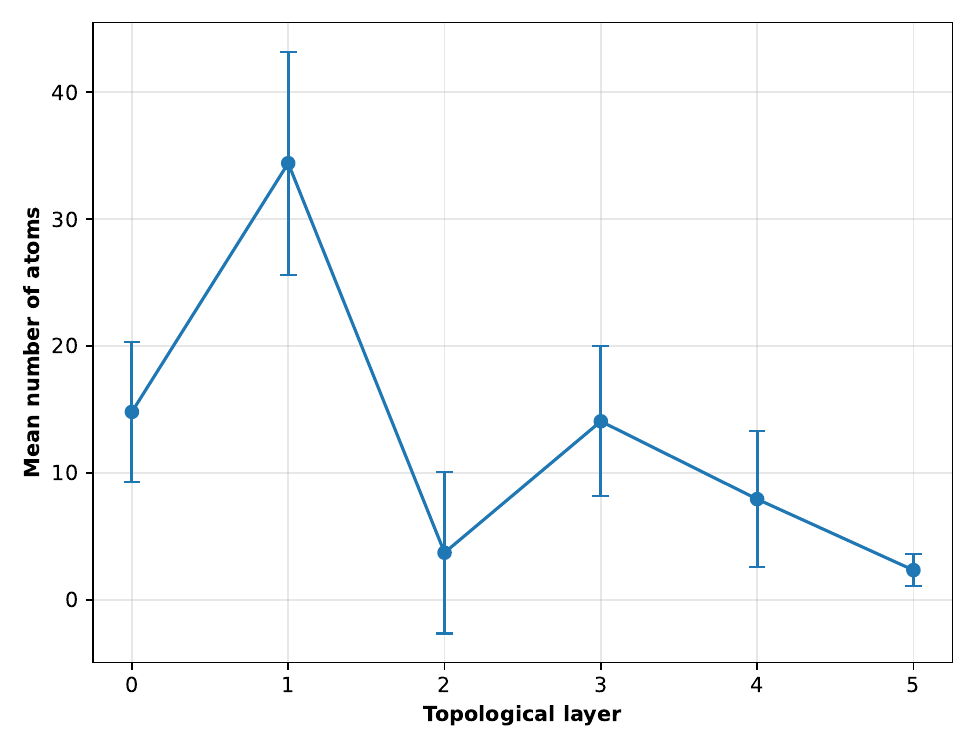}
\caption{Mean atoms per layer.}
\label{fig:mean_atoms}
\end{figure}

\begin{figure}[b!]
\centering
\includegraphics[width=0.75\linewidth]{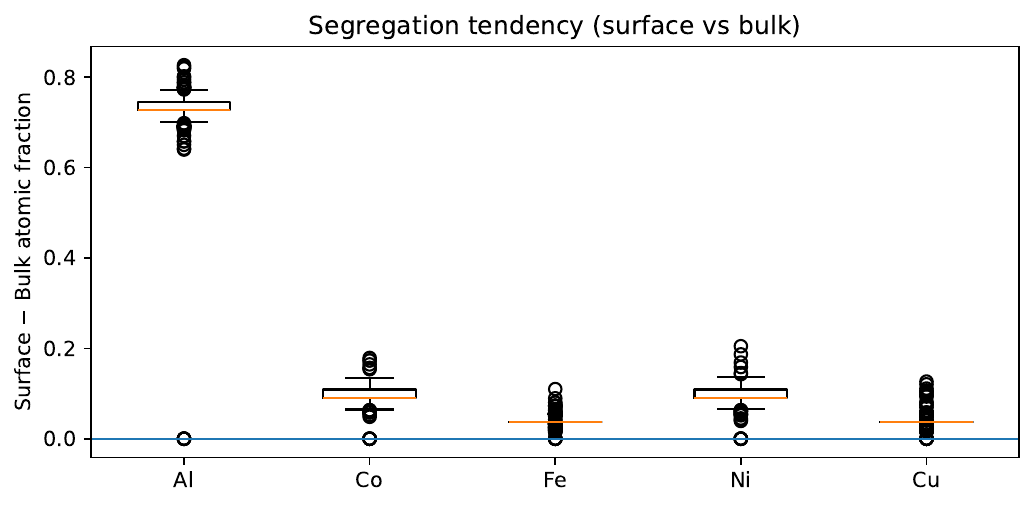}
\caption{Surface--bulk segregation tendencies expressed as the difference between surface
and bulk atomic fractions.
Al exhibits a strong and systematic enrichment at the surface, whereas transition
metal species show substantially weaker segregation.
The consistency of these trends across the dataset highlights the robustness of
layer-dependent chemical heterogeneity and confirms that the descriptor encodes
physically meaningful surface--core contrasts beyond global stoichiometry.}
\label{fig:SA}
\end{figure}

\begin{figure}[b!]
\centering
\includegraphics[width=0.75\linewidth]{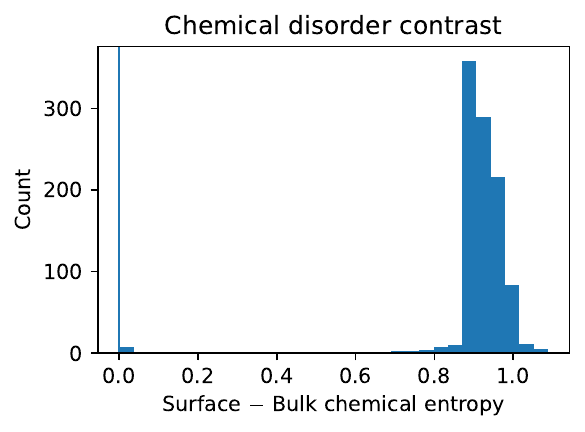}
\caption{Distribution of the chemical entropy contrast between surface and bulk regions,
defined as $\Delta S_{\mathrm{chem}} = S_{\mathrm{chem}}^{\mathrm{surface}} -
S_{\mathrm{chem}}^{\mathrm{bulk}}$, across the full nanoparticle dataset.
Positive values indicate enhanced chemical disorder at the surface relative to the
interior.
The asymmetric distribution, centered at positive $\Delta S_{\mathrm{chem}}$,
demonstrates that surface regions systematically accommodate a higher degree of
chemical mixing, reflecting the combined effects of reduced coordination,
segregation-driven compositional fluctuations, and relaxed bonding constraints.
This result provides quantitative evidence that chemically complex nanoparticles
exhibit intrinsically heterogeneous disorder landscapes, which cannot be captured
by global descriptors alone and are naturally resolved by the proposed
layer-resolved representation.}
\label{fig:SB}
\end{figure}

\begin{figure}[b!]
\centering
\includegraphics[width=\linewidth]{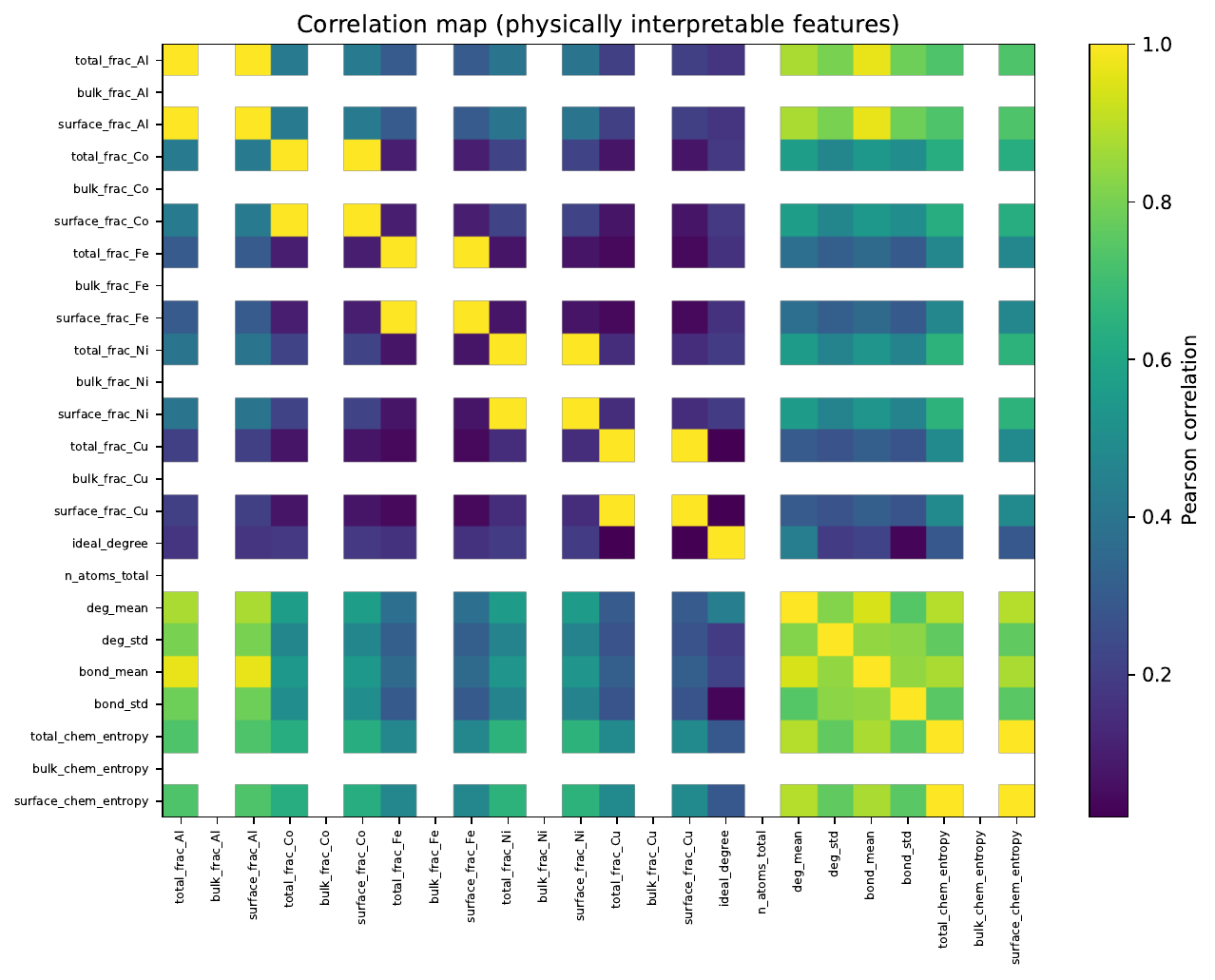}
\caption{Correlation map between selected physically interpretable descriptors,
including global, bulk, and surface-resolved composition, coordination, bonding,
and chemical disorder metrics.
Strong correlations are observed among chemically related quantities (e.g.,
total, bulk, and surface fractions of the same element), reflecting the
constrained global stoichiometry, while cross-correlations between compositional,
topological, and disorder-related features remain moderate.
Notably, chemical entropy descriptors exhibit systematic correlations with
coordination and bond-length statistics, highlighting the coupled nature of
chemical disorder and local topology in chemically complex nanoparticles.
The absence of excessive collinearity across distinct descriptor families
demonstrates that the proposed layer-resolved representation encodes complementary
and physically meaningful information across surface and bulk environments.}
\label{fig:SC}
\end{figure}



